\documentclass[a4paper]{article}

\usepackage[pages=all, color=black, position={current page.south}, placement=bottom, scale=1, opacity=1, vshift=5mm]{background}
\usepackage[margin=1in]{geometry} % full-width

\usepackage{amsmath}
\usepackage{amsthm}
\usepackage{amssymb}

\usepackage[utf8]{inputenc}
\usepackage{hyperref}
\hypersetup{
	unicode,
	pdfauthor={Author One, Author Two, Author Three},
	pdftitle={A simple article template},
	pdfsubject={A simple article template},
	pdfkeywords={article, template, simple},
	pdfproducer={LaTeX},
	pdfcreator={pdflatex}
}

\usepackage[sort&compress,numbers,square]{natbib}
\theoremstyle{plain}

\theoremstyle{definition}

\usepackage[caption=false,font=footnotesize]{subfig}
\usepackage[commentColor=cyan]{algpseudocodex}
\graphicspath{{fig/}}

\usepackage{algorithm}

\algrenewcommand\algorithmicindent{0.8em}%
\algdef{SE}[SUBALG]{Indent}{EndIndent}{}{\algorithmicend\ }%
\algtext*{Indent}
\algtext*{EndIndent}

\usepackage{lipsum}

\title{An algorithm to align a chain of sequences to paths in a pangenome graph}

\author{Görkem Kadir Solun$^1$ \and Ugur Dogrusoz$^1$}

\date{
	$^1$i-Vis Research Lab, Computer Engineering Dept, Bilkent Univ, Ankara, Turkiye 
}

\begin{document}
\maketitle
\begin{abstract}
Affordable, high-quality whole-genome assemblies have made it possible to construct rich pangenomes that capture haplotype diversity across many species. As these datasets grow, they motivate the development of specialized techniques capable of handling the dense sequence variation found in large groups of related genomes. A common strategy is to encode pangenomic information in graph form, which provides a flexible substrate for improving algorithms in areas such as alignment, visualization, and functional analysis. Methods built on these graph models have already shown clear advantages in core bioinformatics workflows, including read mapping, variant discovery, and genotyping.

By integrating multiple sequence and coordinate representations into a single structure, pangenome graphs offer a unified and expressive framework for comparative genomics. Although it remains unclear whether graph-based references will ultimately supplant traditional linear genomes, their versatility ensures that they will play a central role in emerging pangenomic approaches.

This paper introduces an algorithm to mine a chain of sequences in pangenome graphs that might be useful in the functional analysis of pangenome graphs. Specifically, the algorithm calculates all maximal paths in a pangenome graph aligning with a given chain of sequences in the segments of the path vertices, possibly with some maximal gap as specified by the user.\\

\noindent\textbf{Keywords:} pangenome graphs, graph mining, graph algorithms, sequence alignment, ordered chain sequences
\end{abstract}
	
\section{Introduction}

Recent advances in long-read sequencing and genome assembly have enabled the routine generation of high-quality, haplotype-resolved genomes at population scale. These developments have accelerated the adoption of \textit{pangenomes} as a means to represent genomic diversity beyond what is possible with a single linear reference~\cite{Tettelin2005,Medini2008}. Large pangenome projects across diverse organisms have consistently shown that substantial sequence, structural, and haplotypic variation is either poorly represented or entirely absent from traditional linear references, motivating more expressive data models for comparative genomics~\cite{Paten2017,Sherman2019}.

Graph-based representations have emerged as a principled framework for encoding pangenomes, modeling shared and alternative genomic sequences as paths through a common graph structure. In such models, nodes typically represent sequence segments, while edges encode adjacency relationships across assemblies, allowing multiple genomes to be embedded as distinct walks through the same graph~\cite{Paten2017,Novak2017}. This representation unifies multiple coordinate systems into a single substrate and supports a wide range of analyses that are cumbersome to express over linear genomes. As a result, pangenome graphs are increasingly used as a foundational data structure in modern genomics pipelines.

This paper presents an algorithm for identifying chains of sequences in pangenome graphs that are relevant to functional analysis. The proposed method enumerates all maximal paths whose vertex segments align with a specified sequence chain, allowing for user-defined bounds on permissible gaps between consecutive matches.

\section{Method}

We assume that the reader has a working knowledge of standard graph-theoretic concepts and notation~\cite{Diestel2017}.

Let $G=(V,E)$ be a pangenome graph, where $V$ is the set of segments and $E$ contains link, jump, and containment relations between segments. Suppose we are given a chain of ordered sequences of interest $S=(S_1,\dots,S_k)$, where $k\geq 1$. Let $S_i$, $1\leq i\leq k$ denote a segment of length $n_i\geq 1$, where $S_i[j]\in \{A,C,G,T\}$, $1\leq j\leq n_i$. 

\subsection{Idea}
We would like to find all \textit{maximal} paths in $G$ that contain the given chain of sequences (all or a prefix) in the segments of the path vertices, possibly with some gaps. However, we would like to limit these gaps where vertices on the path do not contain any of the segments from the chain by a parameter $l$. In other words, such gaps (i.e., jumps) are limited to a maximum total length of $l\geq 0$. These paths are considered \textit{maximal} in the sense that they cannot be extended to match a longer prefix of the given chain of sequences without exceeding the specified maximum total gap. In addition, we assume that the user may only be interested in paths of a minimum certain match length $1\leq m \leq k$. Furthermore, when a certain segment is reached with varying total gaps, we opt to proceed only with the one with a minimum total gap, pruning any others for efficiency purposes.

\begin{figure}[h!]
\centering{
\includegraphics[width=0.7\linewidth]{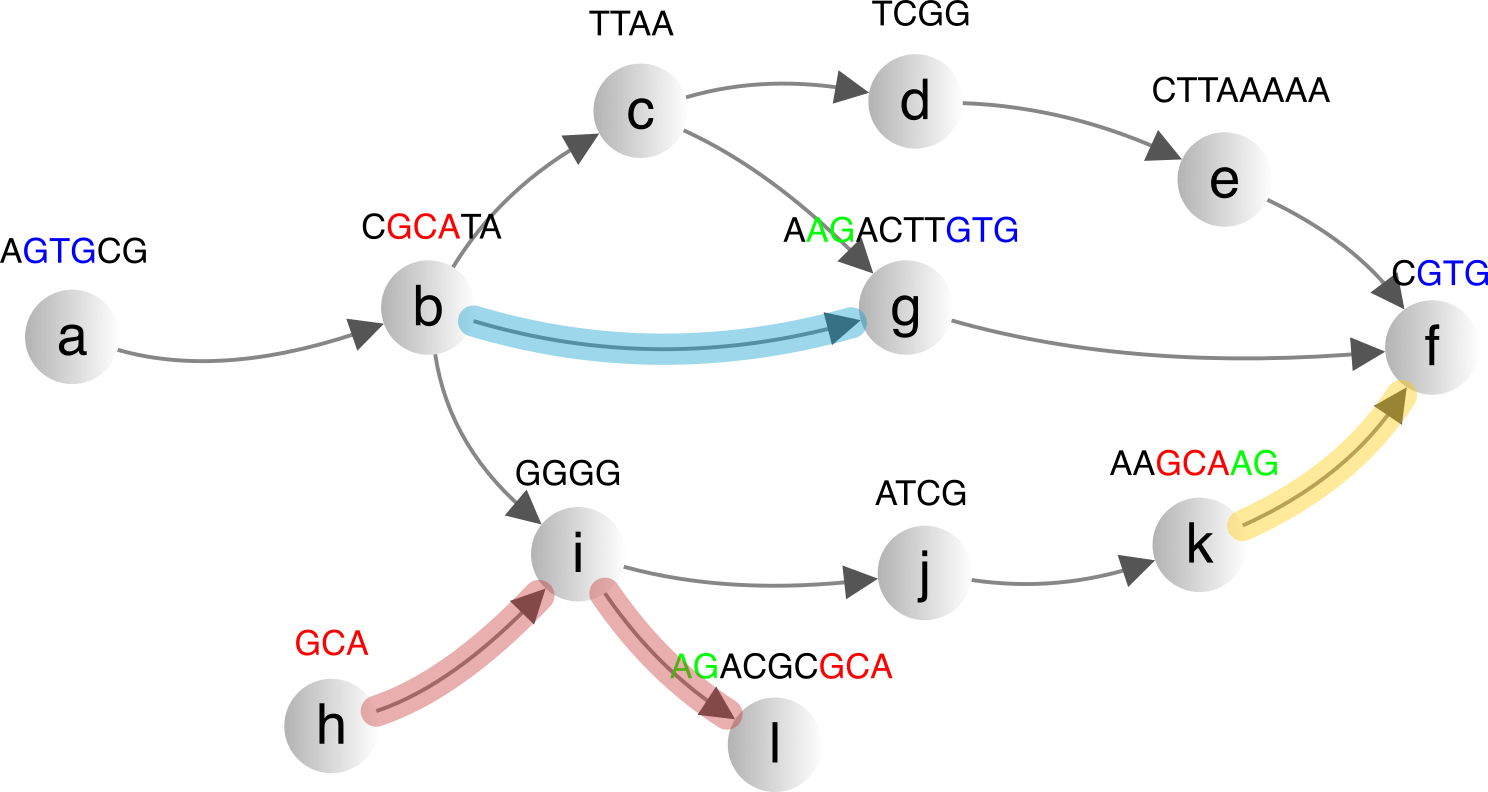}
\caption{A sample pangenome graph $G=(\{a,b,\dots,l\},E)$, where specified sequence chain is $(S_1=\textrm{\color{red}GCA},S_2=\textrm{\color{green}AG},S_3=\textrm{\color{blue}GTG})$, minimum match length $m=2$, and maximal gap $l=1$. Resulting paths are highlighted distinctly.
\label{fig:example}
}}
\end{figure}

Fig.~\ref{fig:example} shows our method with an example. It will start a BFS traversal, seeded at nodes $b$, $h$, $k$, and $l$ matching $S_1$, and will identify these paths:  $(b,(b,g),g)$, $(h,(h,i),i,(i,l),l)$, and $(k,(k,f),f)$ as detailed later. Two of these paths fully match the provided sequence chain and one matches only the first two segments. Notice that $(l)$ is too short to be output, matching just the first sequence. Also note that some paths like $(b,(b,i),i,(i,l),l)$ are pruned in favor of others such as $(h,(h,i),i,(i,l),l)$.

To be more formal, for a given graph $G=(V,E)$ with a specified chain sequence of $S=(S_1,\dots,S_k)$, a maximum gap limit of $l$, and minimum match length of $m$, we would like to calculate:
\[\:\{P=(v_1,\dots,v_{m'})\ |\ 1\leq m'\leq |V|\ \wedge m'\leq k+l\ \wedge \]
\[V'=\{v_{i_1},\dots,v_{i_{k'}}\}\subseteq V\ \wedge \]
\[ P \textrm{ forms a path in } G\ \wedge\ i_{1}=1\ \wedge\ i_{k'}=m'\ \wedge \]
\[ \forall (1\leq j\leq k'\leq k)\ [S_j \in S(v_{i_j})]\ \wedge \]
\[ \forall (1\leq j\leq k'-1)\ [i_j\leq i_{j+1}\ \wedge\ i_{j+1}-i_j \leq l]\:\} \]
where $V'$ is an ordered sequence of vertices in $P$ matching a prefix of ordered sequence $S$ with given constraints, $S(v)$ represents the sequence of the segment node $v$ in $G$ and for sequence strings $S_1$ and $S_2$, $S_1\in S_2$ denotes $S_1$ being a substring of $S_2$.

\subsection{Algorithm}
\begin{algorithm}[t!]
\caption{Find all maximal paths of length at least $m$ and a total gap of at most $l$ that contain a prefix of a specified chain of segments in pangenome graphs}\label{chain-algo}
\begin{algorithmic}[1]
\Function{findSeqChain}{$G=(V, E), S=(S_1,\dots,S_k), m, l$}
\State $V_s \gets \{v\ |\ S_1\in S(v) \}$
\Comment{seed segments}
\State $Q\gets$ \textrm{empty priority queue}
\For{\textbf{each} $v\in V_s$} \Comment{start with all seed nodes}
\State $Q.\Call{insert}{(v,null,(),(),0,0,0)}$
\EndFor

\State $T \gets \emptyset$ \Comment{explored nodes with index of last matched segment}
\State $R \gets \emptyset$ \Comment{resulting path set}

\While{$Q\neq \emptyset$}
\State $(v,e,P,P',i,j,k)\gets Q.\Call{extractMin}$
\If{$(v, i+1)\!\in\!T \wedge i \geq |S|$} \Comment{already visited with a match to $S_{i}$}
\State \textbf{continue}
\EndIf
\State $start\gets S(v).\Call{substring}{j\!+\!1}.\Call{contains}{S_{i+1}}$
\If{$start > 0$} \Comment{$v$ matches next specified sequence}
\If{!($P$ ends with $v$)} \Comment{no previous segment match}
\If{$i\!+\!1\!\geq\!m$} \Comment{at least of a minimum match length}
\State $R\gets R\setminus P$ \Comment{temporarily remove path}
\EndIf
\State $P.\Call{append}{P'}$ \Comment{first append any running gaps}
\If{$e \neq null$} \Comment{$e$ is $null$ for seed nodes}
\State $P.\Call{add}{e}$
\EndIf
\State $P.\Call{add}{v}$ \Comment{$e.target=v$}
\If{$i\!+\!1\geq m$} 
\State $R\gets R \cup P$ \Comment{add (back) extended path}
\EndIf
\EndIf
\State $T \gets T\cup \{(v, i+1)\}$ \Comment{$v$ explored with a match of $S_{i+1}$}
\State $Q.\Call{insert}{(v,e,P,(),i+1,start\!+\!S_i.\Call{length}{},k)}$
\Else \Comment{no match with $S_{i+1}$}
\State $T \gets T\cup \{(v, i)\}$ \Comment{mark $v$ explored with a match of $S_i$}
\If{!($P$ ends with $v$)}
\State $k\gets k\!+\!1$ \Comment{increment total gap}
\EndIf
\If{$k\leq l$} \Comment{prune if gap is too long}
\If{!($P$ ends with $v$)}
\If{$e\neq null$}
\State $P'.\Call{add}{e}$ \Comment{extend the path for the current gap}
\EndIf
\State $P'.\Call{add}{v}$ \Comment{$e.target=v$}
\EndIf
\For{\textbf{each} $w$ with $(v,w)\in E$} \Comment{outgoing neighbors}
\State $Q.\Call{insert}{(w,(v,w),P,P',i,0,k)}$
\EndFor
\EndIf
\EndIf
\EndWhile
\State \textbf{return} $R$
\EndFunction
\end{algorithmic}
\end{algorithm}

One can find all such paths using an algorithm outlined in Algorithm~\ref{chain-algo}, using a customized BFS emanating from all seed nodes containing the first segment in the specified sequence. Here, we use a priority queue of nodes to determine their proper processing order. Each element of this queue consists of 7-tuples $(v,e,P,P',i,j,k)$, where:
\begin{itemize}
    \item $v$ is a node of the pangenome graph that is queued for processing,
    \item $e$ is the edge through which $v$ was reached ($null$ for seed nodes),
    \item $P$ is a path formed by the traversal so far, starting with a seed node and then containing alternating edge and vertex pairs,
    \item $P'$ is a path of the latest gap, formed after the last match, if any (this avoids reporting any matching paths with unnecessary gaps at the \textit{end}),
    \item $i$ keeps the index of the last specified segment $S_i$ that we have matched so far during our traversal,
    \item $j$ marks the end of the consumed part of $v$'s segment $S(v)$ so far, and
    \item $k$ stores the running total gap since the beginning.
\end{itemize} 

We assume characters in a string representing segments are indexed starting with 1, and the following operations on such strings are available:
\begin{itemize}
    \item $\Call{substring}{i}$: returns the substring starting at index $i$,
    \item $\Call{contains}{S_i}$: returns the beginning index of the match with $S_i$, if any, 0 otherwise, and
    \item $\Call{length}$: returns the length of the string.
\end{itemize}

$k$ and $i$ are used as the priority of items in the queue. Lower values of $k$ indicate higher priority. In cases of ties, higher values of $i$ indicate higher priorities.

$\Call{insert}$ and $\Call{extractMin}$ are operations on our priority queue for inserting a new 7-tuple into the queue and extracting the one with the highest priority, respectively. 

Remember that a single node's segment might contain multiple specified segments in a sequence.

Finally, $\Call{add}$ appends the provided vertex or edge to the end of a path. Paths are ordered sequences starting with a vertex, followed by alternating pairs of an edge and a vertex, corresponding to the target of the edge.

In our algorithm, $T$ maintains tuples $(v,i)$, where $v$ is a node and $i$ is an integer, depicting a path reaching vertex $v$ from some seed node, matching the first $i$ sequences in the chain. This is used to avoid redundant traversals. In other words, if vertex $v$ has already been reached, matching the first $i$ sequences in the chain, there is no need to proceed with another similar path as it will not match the given chain any longer than the earlier one. This pruning aims to shorten the algorithm's running time.

\subsection{Time Complexity}
With a careful choice of the associated data structures, the algorithm is expected to run in $O(|V|+|E|+|R|)$ time, where $R$ is a set of calculated paths (i.e., linear in the number of nodes, the number of edges, and the total length of the paths to be reported). Particularly, a hash map could be used to implement explored nodes $T$. Then, each time a node is visited, it can be processed in asymptotically constant running time on average.

%	\paragraph{Acknowledgements} \lipsum[6]
	
%	\newpage
	\bibliography{chain-sequence}
	
\end{document}